\documentclass[prl,twocolumn,showpacs]{revtex4}
\usepackage{epsfig,amsmath}

\begin{document}

\title{Fluctuation-Exchange Study of Antiferromagnetism in Electron-Doped Cuprate Superconductors with Disorder}

\author{Xin-Zhong Yan$^{1,2}$ and C. S. Ting$^1$}
\affiliation{$^{1}$Texas Center for Superconductivity, University of Houston, Houston, TX 77204\\
$^{2}$Institute of Physics, Chinese Academy of Sciences, P.O. Box 603, Beijing 100080, China}

\date{\today}

\begin{abstract}
On the basis of the Hubbard model, we extend the fluctuation-exchange (FLEX) approach to investigating the properties of antiferromagnetic (AF) phase in electron-doped cuprate superconductors. Furthermore, by incorporating the effect of scatterings due to the disordered dopant-atoms into the FLEX formalism, our numerical results show that the antiferromagnetic transition temperature, the onset temperature of pseudogap due to spin fluctuations, the spectral density of the single particle near the Fermi surface, and the staggered magnetization in the AF phase as a function of electron doping can consistently account for the experimental measurements. 
\end{abstract}

\pacs{75.10.-b, 74.25.Ha, 71.10.-w, 71.10.Fd} 

\maketitle

%\vfill

The cuprate high-temperature superconductors are the typical quasi-two-dimensional (Q2D) strongly-correlated electron systems. At low temperature, the electron-doped cuprates (EDC) such as Nd$_{2-x}$Ce$_x$CuO$_4$ (NCCO) and Pr$_{2-x}$Ce$_x$CuO$_4$ (PCCO) are in the antiferromagnetic (AF) state within a wide doping range \cite{Luke,Kubo,Mang,Thurston}. The Fermi surface (FS) evolution with electron doping observed by the angle-resolved photoemission spectroscopy experiment \cite{Armitage,Damascelli} reveals the microscopic information about the antiferromagnetism. For understanding the physics of FS evolution in EDC, much of the works have been carried out by using either the Hubbard or $t-J$ model \cite{Kusko,Kusunose,Senechal,Markiewicz,Kyung,Yuan,Yan}. But there are difficulties with the existing theories in describing the AF phase transition and the FS evolution in a consistent manner. To explain the FS evolution and the staggered magnetization as a function of doping, most of the theories have to assume a doping-dependent Hubbard $U$. Since $U$ is the on-site Coulomb interaction, it is hard to understand how it could be screened by conduction electrons. Though the mean-filed theory (MFT) yields qualitative explanation for FS evolution \cite{Kusko,Yuan,Yan}, the predicted AF transition temperature $T_N$ is well known to be too high \cite{Markiewicz,Yan}. In order to reduce the magnitude of $T_N$, the self-consistent random-phase approximation which takes into account the effect of spin fluctuations has been employed, but the obtained $T_N$ \cite{Markiewicz} is  a non-monotonic function of doping concentration $x$, contrary to the experimental observations. Therefore, a quantitative understanding of the AF phase transition in EDC remains to be a challenging issue which so far has not been properly addressed.

In this work, we extend the fluctuation-exchange (FLEX) approach based on the Hubbard model \cite{Bickers}, which was designed for studying the spin fluctuations in a two-dimensional strongly-correlated electron system, to investigating the antiferromagnetism below $T_N$ in NCCO or PCCO. Furthermore, by incorporating the effect of scatterings due to the disordered dopant-atoms Ce into this formalism, we are able to show that the numerically obtained $T_N$, the onset temperature of pseudogap due to spin fluctuations, the spectral density of the single particle near FS, and the staggered magnetization $m$ in the AF phase as a function of $x$ are all in good agreement with the relevant experimental measurements. Here $U$ is doping independent and each dopant Ce atom is expected to release one electron in the CuO planes. With increasing doping, the scattering effect due to dopant atoms becomes more pronounced which is expected to expedite the suppression of the AF order.

For describing the electron system, we use the Hubbard model including the impurity field:
\begin{equation}
H= \sum_{\vec k,\sigma}\xi_{\vec k} c_{k\sigma}^{\dagger}c_{k\sigma} 
+ \frac{U}{N} \sum_{\vec q}n_{\vec q\uparrow} n_{-\vec q\downarrow}
+ \sum_{i}v_{i}n_{i}
\label{Hubbard}
\end{equation}
where $\xi_{\vec k} = \epsilon_{\vec k}-\mu$ with $\epsilon_{\vec k}$ as the tight-binding energy, $\mu$ the chemical potential, $c_{\vec k\sigma}^\dagger$ ($c_{\vec k\sigma}$) is the electron creation (annihilation) operator of spin-$\sigma (= \pm 1$ for up and down spins, respectively) in state of momentum $\vec k$, $n_{q\sigma}=\sum_{\vec k}c_{\vec k\sigma}^\dagger c_{\vec k+\vec q \sigma}$, and $N$ is the total number of the lattice sites. $v_i$ in the last term is the impurity potential acting on the electron with density $n_{i}$ at position $i$. In NCCO or PCCO, the concentration $N_i$ of the dopant Ce atoms is the same as the electron doping concentration $x$. The electron system under consideration is the Q2D one with weak interlayer coupling. 

We begin to work with the Green's function $G$. The formalism for determining $G$ is to find out the self-energy $\Sigma$ as a functional of $G$. In the normal state, following the FLEX approximation \cite{Bickers}, we can easily get the self-energy functional in the presence of impurities. $\Sigma$ is diagrammatically shown in Fig. 1. The diagrams in Fig. 1(a) come from the Hartree term. The first one includes the diffuson correction due to the particle-hole propagator under the impurity scattering. The second one contains the Cooperon correction. Figures 1(b) and 1(c) are the FLEX diagrams with the decoration of the impurity scattering. The ring diagrams in Fig. 1(b) represent the contributions from the longitudinal spin and density fluctuations. The transverse spin fluctuations are taken into account in Fig. 1(c). The impurity insertions in the above diagrams are illustrated in Figs. 1(d) and 1(e). Such insertions ensure the spherical symmetry for the longitudinal and transverse spin susceptibilities in the presence of impurity scatterings in the normal state. The process of the single electron scattering off impurities is shown in Fig. 1(f). Since the antiferromagnetism below $T_N$ is also under consideration, instead of writing down the formula for Fig. 1, we need to extend the normal-state expressions to include the AF order below $T_N$.

\begin{figure}
\centerline{\epsfig{file=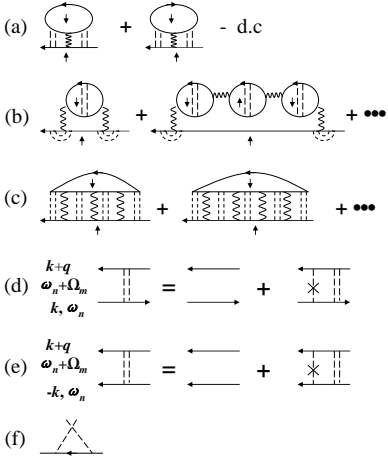,width=5.5 cm}}
%\vskip 2mm
\caption{Diagrams for the self-energy in normal state. The wavy line represents the on-site Hubbard $U$. The dashed single line with a cross denotes the impurity scattering. (a) Lowest order (Hartree) terms, where a double counting (d.c.) without the impurity scattering should be subtracted, (b) ring diagrams representing the longitudinal spin and density fluctuations, (c) diagrams for transverse spin fluctuations, (d) vertex corrections due to the impurity scattering for the particle-hole and (e) particle-particle channels, (f) self-energy due to the impurity scattering.}
\end{figure}
 
The AF state is characterized by the staggered magnetization $m$ with commensurate wavevector $\vec Q$. In the AF state, the operator $n_{\vec Q,\sigma} \to -\sigma Nm/2$ is a macroscopic quantity. The term of $\vec q = \vec Q$ in the Hubbard interaction in Eq. (\ref{Hubbard}) should be singled out, and be treated as
\begin{equation}
\frac{U}{N}n_{\vec Q\uparrow} n_{-\vec Q\downarrow}\approx \Delta\sum_{\vec k}(c_{\vec k\uparrow}^\dagger c_{\vec k+\vec Q \uparrow}-c_{\vec k\downarrow}^\dagger c_{\vec k+\vec Q \downarrow})  
\label{MFT}
\end{equation}
where $\Delta = Um/2$ is the order parameter. Though Eq. (\ref{MFT}) seems like the MFT, we here also take into account simultaneously the fluctuation terms $\vec q \ne \vec Q$ in the interaction. Because of Eq. (\ref{MFT}), we need to use the Green's function $\hat G\equiv -\langle T_{\tau}\psi_{\vec k\sigma}(\tau)\psi^{\dagger}_{\vec k\sigma}(\tau')\rangle$ defined as a $2\times 2$ matrix with $\psi^{\dagger}_{\vec k\sigma}\equiv (c^{\dagger}_{\vec k\sigma},\sigma c^{\dagger}_{\vec k+\vec Q \sigma})$. Since the potential ($\pm\Delta$) in Eq. (\ref{MFT}) for opposite spin electrons has opposite sign, $\hat G$ so defined is independent of the spin. The order parameter $\Delta$ can then be written as
\begin{equation}
\Delta=-\frac{U}{N\beta}\sum_{k}G_{12}(k) \label{op}
\end{equation}
where $\beta$ is the inverse of the temperature $T$, and $k \equiv (\vec k, i\omega_n)$ with $\omega_n$ the electronic Matsubara frequency. From Eq. (\ref{MFT}) with the impurity influence, the off-diagonal self-energy $\Sigma_{12}(k)$ is obtained as in Fig. 2. 

\begin{figure}
\centerline{\epsfig{file=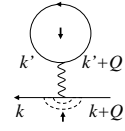,width=1.7 cm}}
\caption{Off-diagonal self-energy $\Sigma_{12}$.}
\end{figure}

To extend the normal-state self-energy shown in Fig. 1 to that of the AF state, we note that the transverse spin fluctuation in Fig. 1(c) should include the collective modes, especially the Goldstone mode (GM). The latter is consistent with the gap equation (\ref{op}). This consistency gives rise to a constraint on the form of $\Sigma_{12}$, and one can show that Fig. 2 is the only graph for $\Sigma_{12}$ within the FLEX approximation. In addition, the insertion of impurity scatterings in the transverse spin fluctuation is between the opposite spin electrons, while in Fig. 2 it is between the same spin electrons. In order to ensure the consistency between GM and Eq. (\ref{op}), these insertions should be the same; so the processes due to $G_{12}$ should be excluded. Keeping these points in mind, the expressions for the self-energy in the AF state can be obtained.

To give an explicit expression for the self-energy, we start with the vertex corrections shown in Figs. 1(d) and 1(e). They can be expressed by the same formula,
\begin{equation}
\Gamma(q,i\omega_n) = [1-N_iv_i^2X(q,i\omega_n)]^{-1} \label{vc}
\end{equation}
with
$X(q,i\omega_n) = \frac{1}{N}\sum_{\vec k}G_{11}({\vec k},i\omega_n)G_{11}({\vec k}+{\vec q},i\omega_n+i\Omega_m)$,  where $q = ({\vec q},i\Omega_m)$ with $\Omega_m$ the bosonic Matsubara frequency. With the vertex correction, the logitudinal $\chi_z$ and transverse $\chi^t$ spin susceptibilities are given by
\begin{eqnarray}
\chi_z(q) = \frac{1}{\beta}\sum_n\Gamma(q,i\omega_n)X_{+}(q,i\omega_n), \label{chiz}\\
\chi^t_{11}(q) =\frac{1}{\beta}\sum_n\Gamma(q,i\omega_n)X_{-}(q,i\omega_n), \label{ch11}\\
\chi^t_{12}(q) = \frac{1}{\beta}\sum_n\Gamma(q,i\omega_n)X_o(q,i\omega_n)\Gamma(q+Q,i\omega_n), \label{ch12}
\end{eqnarray}
with
\begin{eqnarray}
X_{\pm}(q,i\omega_n) = \frac{1}{N}\sum_{\vec k}[G_{11}(k)G_{11}(k+q)\pm G_{12}(k)G_{12}(k+q)] \nonumber\\ 
X_{o}(q,i\omega_n)=\frac{1}{N}\sum_{\vec k}[G_{11}(k)G_{12}(k+q)-G_{12}(k)G_{11}(k+q)] \nonumber
\end{eqnarray}
and $Q$ means $Q \equiv (\vec Q,0)$. $\chi^t_{11}(q)$ is the susceptibility between the transverse spin fluctuations of the same momentum $q$, while $\chi^t_{12}(q)$ between that of the momenta $q$ and $q+Q$. For $\chi^t_{21}$ and $\chi^t_{22}$, we have $\chi^t_{21}(q)=\chi^t_{12}(q)$ and $\chi^t_{22}(q)= \chi^t_{11}(q+Q)$. 
The expression for $\Sigma_{11}$ reads, 
\begin{eqnarray}
\Sigma_{11}(k)&=&-\frac{2U}{N\beta}\sum_{q}G_{11}(k+q)[\Gamma^2(q,i\omega_n)-1]+ \Sigma_i(i\omega_n) \cr\cr
& &-\frac{1}{N\beta}\sum_{q}G_{11}(k+q)V_{\rm eff}(q)\Gamma^2(q,i\omega_n) , \label{sef}
\end{eqnarray}
with $V_{\rm eff}(q) = V_z(q) + V_{11}(q)$ and
\begin{eqnarray}
V_z(q) = \frac{1}{2}\frac{U^2\chi_z(q)}{1+U\chi_z(q)}
+\frac{1}{2}\frac{U^2\chi_z(q)}{1-U\chi_z(q)}, \label{vz}\\
V_{11}(q) = -U\{[1+U\hat\chi^t(q)]^{-1} - 1+U\hat\chi^t(q)\}_{11}, \label{v11}\\
\Sigma_i(i\omega_n) = \frac{N_iv_i^2}{N}\sum\limits_{\vec k}G_{11}({\vec k},i\omega_n).  \label{tm}
\end{eqnarray}
The first term in right-hand side of Eq. (\ref{sef}) comes from the Hartree terms in Fig. 1(a). The contribution to the self-energy in the absence of impurity scattering is subtracted because of a double counting and the remaining constant absorbed in the chemical potential. $\Sigma_i$ corresponds to Fig. 1(f). The last term is due to the density and spin fluctuations. At $q=Q$, $V_{11}$ is divergent because of $1+U\chi^t_{11}(Q)=0$, which means the existance of GM and is consistent with Eq. (\ref{op}). In the normal state with $G_{12}=0$, if $v_i = 0$, Eq. (\ref{sef}) reduces to the FLEX theory \cite{Pao,Monthoux}. The off-diagonal self-energy according to Fig. 2 can be written as
\begin{equation}
\Sigma_{12}(k)=\Delta\Gamma(Q,i\omega_n). \label{off}
\end{equation}
The other elements of $\hat\Sigma(k)$ are given by $\Sigma_{21}(k)=\Sigma_{12}(k)$, and $\Sigma_{22}(k)=\Sigma_{11}(k+Q)$.

We here consider the interlayer coupling (IC). For NCCO or PCCO in which the CuO layers are staggered stacking, the AF spin coupling between the nearest layers is determined by the local lattice distortions or any other asymmetry; the ideal stacking leads to zero spin-coupling. We here will not consider this complicated situation in detail, but artificially introduce a weak IC in the following manner. Note that IC becomes significant only at the singularity in Eq. (\ref{v11}) where $1+U\chi^t_{11}(\vec Q,0)=0$ with $\vec Q \equiv (\pi,\pi,Q_z)$. At $\vec q \to \vec Q$, we should have $1+U\chi^t_{11}(\vec q,0) \approx |\vec c\cdot(\vec q - \vec Q)|^2$, where $\vec c$ is a constant vector. The IC is reflected by the small $z$-component constant $c_z$. This constant controls the overall magnitude of $T_N$. However, the shape of the AF phase boundary should be determined predominantly by the intra-layer coupling. Noting that $\hat G$ and $\chi$'s very weakly depend on the weak IC, we then approximate them as the two-dimensional functions by adding a $q_z$-dependent term $c_z^2(q_z-Q_z)^2$ to the denominator of the final expression of $\{[1+U\chi^t(q)]^{-1}\}_{11}$ in the first term of Eq. (\ref{v11}) and also to $1+U\chi_z(q)$ (which vanishes at $T_N$) in the first term of Eq. (\ref{vz}). From our numerical calculation, the magnitude of in-plane component of $\vec c$ is found to be near the order of unity. We then choose $c^2_z = 10^{-3}$ in our calculation. Since the $q_z$-integral is over a period with $Q_z$ centered, the final result is independent of $Q_z$.

For studying the AF properties in EDC, in our numerical calculation, the parameters in unit of $t = 1$ [the nearest-neighbor (NN) hopping] in the model are chosen as $t' = -0.25$ (next NN hopping), $t'' = 0.1$ (third NN hopping), $U = 8$, and $v_i = 2$. The parameters $t'$, $t''$ and $U$, are the typical ones used in the literatures \cite{Senechal}. With the recently developed algorithms for dealing with the summation over the Matsubara frequency and the Fourier transform of effective interaction with long-wavelength singularity \cite{Yan2}, we have numerically obtained the self-consistent solution to the Green's function.

\begin{figure}
\centerline{\epsfig{file=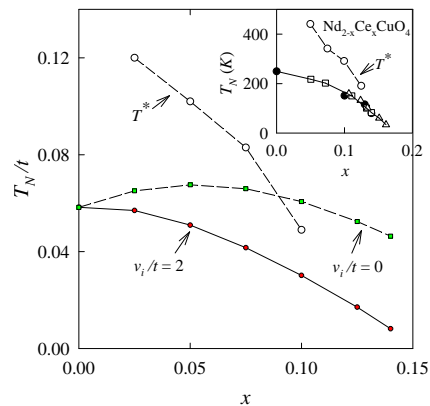,width=5.5 cm}}
\caption{Transition temperature $T_N$ as function of doping concentration $x$. The circles are the onset temperatures $T^{*}$ of the pseudogap. The solid circles \cite{Luke}, triangles \cite{Kubo}, squares \cite{Mang}, and the circles \cite{Onose} in the inset are the experimental data. The lines are for the eye.}
\end{figure}

In Fig. 3, the numerical result for $T_N$ is presented as a function of $x$. The curve of $v_i = 0$ corresponds to the FLEX calculation. For comparison, the experimental data for NCCO are presented in the inset \cite{Luke,Kubo,Mang}. It should be mentioned that the present result for $T_N$ is one order of magnitude smaller than that of MFT \cite{Markiewicz,Yan}. Even though the FLEX gives the overall lower $T_N$, its nonmonotonic behavior does not reflect the feature of the experimental data. In contrast, the calculation with impurity strength $v_i =2$ can produce a monotonic decreasing function $T_N$ of $x$, in fairly good agreement with the experiment if $t$ is choosing to be 0.38 eV. This result stems from a combination effect of spin fluctuations and impurity scatterings. For comparison, we also depict our result on the onset temperature $T^{*}$ of the pseudogap which is originated in the spin fluctuations and exhibits itself as a  depression in the density of states around the chemical potential. Our calculated $T^{*}$ is consistent with the experimental data extracted from measurements on optical conductance \cite{Onose}. Similar result on $T^{*}$ has also been obtained by Kyung {\it et al.} \cite{Kyung} using a different approach with doping dependent $U$. 

\begin{figure}
\centerline{\epsfig{file=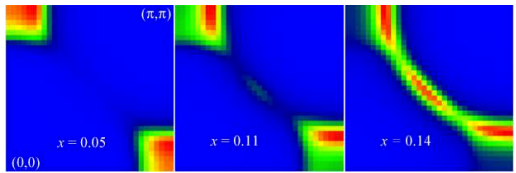,width=6.5 cm}}
\caption{(Color online) Occupied spectral density around the chemical potential at doping concentrations $x = 0.05$, $x = 0.11$, and $x = 0.14$, and at $T/t = 0.003$.}
\end{figure}

Shown in Fig. 4 are the occupied spectral densities $-f(E){\rm Im}G_{11}(\vec k,E+i0^+)/\pi$ (with $f$ the Fermi distribution function) at the chemical potential integrated within the energy window of $-0.05 t < E < 0.05 t$ for three doping concentrations at $T/t = 0.003$. In the AF phase, the original single band folds into a lower and an upper Hubbard bands. At $x = 0.05$, there exists extended gap from $(0,\pi)$ to $(\pi,0)$. At $x = 0.11$ and 0.14, gaps show up around the hot spots. For larger doping where the AF phase diminishes and the spectrum becomes a single curve. This result essentially illustrates the FS evolution with $x$, which has also been qualitatively explained by the two-band picture in the MFT \cite{Kusko,Yan,Yuan}. As seen from Fig. 4, the feature of the experimental result is well reproduced by the present calculation. 

Figure 5 exhibits the staggered magnetization $m$ as a function of $x$ at $T = 0 K$ (obtained by extrapolation from the results of finite temperatures) and the comparison with the experimental data \cite{Mang,Rosseinsky}. The experimental value for $m(0)$ at Cu$^{2+}$ is about 0.5 in NCCO and $\sim 0.4$ in PCCO \cite{Thurston}, which are comparable with the present result 0.44. The behavior of $m$ obtained by the present calculation is in very good agreement with the experiments. Kusko {\it et al.} \cite{Kusko} using the MFT with a doping-dependent $U$ have obtained $m(x)$ which is a linearly decreasing function of $x$ with  $m=0$ at $x \approx 0.14$ and $m(0)\approx 0.4$. This result agrees well with the experimental measurements except near the doping region where $m$ goes to zero. Though the dynamical-MFT calculation by S\'en\'echal {\it et al.} \cite{Senechal} yields a reasonable behavior for $m(x)/m(0)$, $m(0) \approx 0.7$ is too large. 

\begin{figure}
\centerline{\epsfig{file=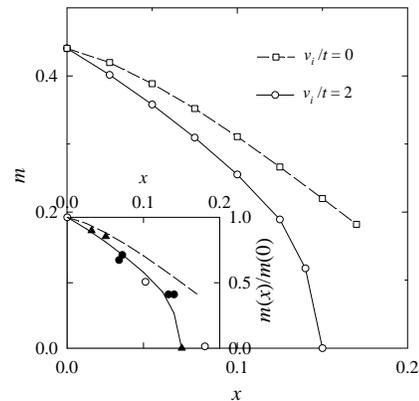,width=5.5 cm}}
\caption{Staggered magnetization $m$ as function of $x$ at $T = 0 K$. The inset shows experimental data (solid and hollow circles \cite{Mang}, solid triangles \cite{Rosseinsky}) compared with the present calculation.}
\end{figure}

In summary, we have extended the FLEX approach to the investigation of the AF phase in electron-doped cuprates. A self-consistent treatment of the effect of spin fluctuations in combination with scatterings of electrons due to the disordered dopant atoms lead to the suppression of both $m$ and $T_N$. The present approach provides numerical results which could consistently account for the experimental measurements on the doping dependences of the Fermi surface evolution and the staggered magnetization $m$ at low temperature, as well as the AF transition temperature $T_N$ and the onset temperature $T^{*}$ for pseudogap formation.  

The authors wish to thank Dr. Q. Yuan for useful discussions. This work was supported by a grant from the Robert A. Welch Foundation under No. E-1146, the Texas Center for Superconductivity at University of Houston, and the National Basic Research 973 Program of China under grant number 2005CB623602.

\end{document}